\documentclass[12pt]{article}
\usepackage{amssymb,amscd,array}
\catcode `\@=11
\@addtoreset{equation}{section}

\newtheorem{thm}{Theorem}[section]

\def\qed{\blacksquare}
\newcommand{\be}{\begin{equation}}
\newcommand{\ee}{\end{equation}}
\newcommand{\bea}{\begin{eqnarray}}
\newcommand{\eea}{\end{eqnarray}}
\begin{document}
\begin{titlepage}

\begin{center}
{\bf \Large{Massive Yang-Mills Fields in Interaction with Gravity\\}}
\end{center}
\vskip 1.0truecm
\centerline{D. R. Grigore, 
\footnote{e-mail: grigore@theory.nipne.ro}}
\vskip5mm
\centerline{Department of Theoretical Physics, Institute for Physics and Nuclear
Engineering ``Horia Hulubei",}
\centerline{Institute of Atomic Physics}
\centerline{Bucharest-M\u agurele, P. O. Box MG 6, ROM\^ANIA}
\vskip 1cm
\centerline{G. Scharf
\footnote{e-mail: scharf@physik.unizh.ch}}
\vskip5mm
\centerline{Institut f\"ur Theoretische Physik, Universit\"at Z\"urich,} 
\centerline{Winterthurerstr. 190 , CH-8057 Z\"urich, SWITZERLAND}

\vskip 2cm
\bigskip \nopagebreak
\begin{abstract}
\noindent
We determine the most general form of the interaction between the gravitational field and an arbitrary Yang-Mills system of fields (massless and massive).  We work in the perturbative quantum framework of the causal approach (of Epstein and Glaser) and use a cohomological definition of gauge invariance for both gauge fields. We also consider the case of massive gravity. We discuss the question whether gravity couples to the unphysical degrees of freedom in the Yang-Mills fields.
\end{abstract}
\end{titlepage}

\section{Introduction}

The general framework of perturbation theory consists in the construction of the chronological products such that Bogoliubov axioms are verified \cite{BS}, \cite{DF}, \cite{EG}; for every set of Wick monomials 
$ 
W_{1}(x_{1}),\dots,W_{n}(x_{n}) 
$
acting in some Fock space 
$
{\cal H}
$
generated by the free fields of the model one associates the operator
$ 
T_{W_{1},\dots,W_{n}}(x_{1},\dots,x_{n}); 
$  
all these expressions are in fact distribution-valued operators called chronological products. Sometimes it is convenient to use another notation: 
$ 
T(W_{1}(x_{1}),\dots,W_{n}(x_{n})). 
$ 
The construction of the chronological products can be done recursively according to Epstein-Glaser prescription \cite{EG}, \cite{Gl} (which reduces the induction procedure to a distribution splitting of some distributions with causal support). These products are not uniquely defined but there are some natural limitation on the arbitrariness. If the arbitrariness does not grow with $n$ we have a renormalizable theory.

Gauge theories describe particles of higher spin. Usually such theories are not renormalizable. However, one can save renormalizability using ghost fields. Such theories are defined in a Fock space
$
{\cal H}
$
with indefinite metric, generated by physical and unphysical fields (called {\it ghost fields}). One selects the physical states assuming the existence of an operator $Q$ called {\it gauge charge} which verifies
$
Q^{2} = 0
$
and such that the {\it physical Hilbert space} is by definition
$
{\cal H}_{\rm phys} \equiv Ker(Q)/Im(Q).
$
The space
$
{\cal H}
$
is endowed with a grading (usually called {\it ghost number}) and by construction the gauge charge is raising the ghost number of a state. Moreover, the space of Wick monomials in
$
{\cal H}
$
is also endowed with a grading which follows by assigning a ghost number to every one of the free fields generating
$
{\cal H}.
$
The graded commutator
$
d_{Q}
$
of the gauge charge with any operator $A$ of fixed ghost number
\be
d_{Q}A = [Q,A]
\ee
is raising the ghost number by a unit. It means that
$
d_{Q}
$
is a co-chain operator in the space of Wick polynomials. From now on
$
[\cdot,\cdot]
$
denotes the graded commutator.
 
In a gauge theory one assumes also that there exists a Wick polynomial of null ghost number
$
T(x)
$
called {\it the interaction Lagrangian} such that
\be
~[Q, T] = i \partial_{\mu}T^{\mu}
\ee
for some other Wick polynomials
$
T^{\mu}.
$
This relation means that the expression $T$ leaves invariant the physical states, at least in the adiabatic limit. In all known models one finds out that there exists a chain of Wick polynomials
$
T^{\mu},~T^{\mu\nu},~T^{\mu\nu\rho},\dots
$
such that:
\be
~[Q, T] = i \partial_{\mu}T^{\mu}, \quad
[Q, T^{\mu}] = i \partial_{\nu}T^{\mu\nu}, \quad
[Q, T^{\mu\nu}] = i \partial_{\rho}T^{\mu\nu\rho},\dots
\label{descent}
\ee
In all cases
$
T^{\mu\nu},~T^{\mu\nu\rho},\dots
$
are completely antisymmetric in all indices; it follows that the chain of relation stops at the step $4$ (if we work in four dimensions). We can also use a compact notation
$
T^{I}
$
where $I$ is a collection of indices
$
I = [\nu_{1},\dots,\nu_{p}]~(p = 0,1,\dots,)
$
and the brackets emphasize the complete antisymmetry in these indices. All these polynomials have the same canonical dimension
\be
\omega(T^{I}) = \omega_{0},~\forall I
\ee
and because the ghost number of
$
T \equiv T^{\emptyset}
$
is supposed null, we then also have:
\be
gh(T^{I}) = |I|.
\ee
One can write compactly the relations (\ref{descent}) as follows:
\be
d_{Q}T^{I} = i~\partial_{\mu}T^{I\mu}.
\label{descent1}
\ee

For concrete models the equations (\ref{descent}) can stop earlier: for instance in the case of gravity
$
T^{\mu\nu\rho\sigma} = 0.
$

Cohomology problems of the type (\ref{descent}) have been extensively studied in the more popular approach to quantum gauge theory based on functional methods (following from some path integration method) \cite{Dr} (and refs. given there). In this setting the co-chain operator is non-linear and makes sense only for classical field theories. On the contrary, in the causal approach the co-chain operator is linear so the cohomology problem makes sense directly in the Hilbert space of the model. For technical reasons one needs however a classical field theory machinery to analyze the descent equations more easily.

In some previous papers \cite{cohomology}, \cite{cohomology2} one of us has developed a machinery of analyzing systematically the descent equations. Here we want to apply these methods for the interaction between gravitation and Yang-Mills fields. We consider the case of massless and massive gravity. The case of massless Yang-Mills fields was analyzed in \cite{cohomology2} and here we provide the general case. We will use a convenient geometric setting for our problem presented in \cite{cohomology}.

In the next Section we introduce the gauge structure on the various free quantum fields. In Section \ref{q} we remind the reader the cohomology of the operator 
$
d_{Q}
$
for Yang-Mills models and gravitation. Using this cohomology and the algebraic Poincar\'e lemma we can solve the descent equations in Section \ref{int1} and we determine the interaction between massless (or massive) gravity and Yang-Mills fields in the most general case.

\section{Free Fields of Spin $1$ and $2$\label{free}}
We remind here some results and notations from \cite{cohomology} and \cite{cohomology2}. The Hilbert space 
$
{\cal H}.
$
we use is of Fock type and it should describe particles of spin $1$ and $2$ with null or positive mass; we will denote by
$\Omega$ 
the vacuum state in
$
{\cal H}.
$
The Pauli-Jordan distribution of mass $m$ is denoted by
$
D_{m}
$
and
$
D_{m}^{(+)}
$
is its positive frequency part. The Minkowski metrics (with diagonal $1, -1, -1, -1$) is denoted by
$
\eta_{\mu\nu}.
$
We will always mean by
$
[\cdot,\cdot]
$
the graded commutator.
The Hilbert space is generated by Bose and Fermi fields so it is useful to introduce a grading in 
$
{\cal H}
$
as follows: every state which is generated by an even (odd) number of Fermi fields and an arbitrary number of Bose fields is even (resp. odd).
\subsection{Massless Vector Fields\label{s=1,m=0}}

We consider a vector space 
$
{\cal H}
$
of Fock type generated (in the sense of Botchers' theorem) by the vector field 
$
A_{\mu}
$ 
(with Bose statistics) and the scalar fields 
$
u, \tilde{u}
$
(with Fermi statistics). The Fermi fields are usually called {\it ghost fields} and one can introduce in a natural way a ghost number operator in the Fock space ($u$ has ghost number $1$ and $\tilde{u}$ has ghost number $- 1$). We suppose that all these (quantum) fields are of null mass. In this vector space we can define a masculine form 
$<\cdot,\cdot>$
in the following way: the (non-zero) $2$-point functions are by definition:
\bea
<\Omega, A_{\mu}(x_{1}) A_{\mu}(x_{2})\Omega> =i~\eta_{\mu\nu}~D_{0}^{(+)}(x_{1} - x_{2}),
\nonumber \\
<\Omega, u(x_{1}) \tilde{u}(x_{2})\Omega> =- i~D_{0}^{(+)}(x_{1} - x_{2})
\qquad
<\Omega, \tilde{u}(x_{1}) u(x_{2})\Omega> =- i~D_{0}^{(+)}(x_{1} - x_{2})
\eea
and the $n$-point functions are generated according to Wick theorem. To extend the masculine form to
$
{\cal H}
$
we define the conjugation by
\be
A_{\mu}^{\dagger} = A_{\mu}, \qquad 
u^{\dagger} = u, \qquad
\tilde{u}^{\dagger} = - \tilde{u}.
\ee

Now we can introduce the operator $Q$ according to the following formulas:
\bea
~[Q, A_{\mu}] = i~\partial_{\mu}u,\qquad
[Q, u] = 0,\qquad
[Q, \tilde{u}] = - i~\partial_{\mu}A^{\mu}
\nonumber \\
Q\Omega = 0
\label{Q-0-1}
\eea
where by 
$
[\cdot,\cdot]
$
we mean the graded commutator. One can prove that $Q$ is well defined. Indeed, we have the causal commutation relations 
\be
~[A_{\mu}(x_{1}), A_{\mu}(x_{2}) ] =i~\eta_{\mu\nu}~D_{0}(x_{1} - x_{2})~\cdot I,
\qquad
[u(x_{1}), \tilde{u}(x_{2})] = - i~D_{0}(x_{1} - x_{2})~\cdot I
\ee
and the other commutators are null. The operator $Q$ should leave invariant these relations, in particular 
\be
[Q, [ A_{\mu}(x_{1}),\tilde{u}(x_{2})]] + {\rm cyclic~permutations} = 0
\ee
which is true according to the previous definition. The usefullness of this construction follows from:
\begin{thm}
The operator $Q$ verifies
$
Q^{2} = 0.
$ 
The factor space
$
Ker(Q)/Ran(Q)
$
is isomorphic to the Fock space of particles of zero mass and helicity $1$ (photons, gluons). 
\end{thm}

$Q$ is usually called {\it gauge charge operator}, the relations (\ref{Q-0-1}) define the gauge structure of the free fields. We notice that the operator $Q$ raises the ghost number of a state (of fixed ghost number) by an unit.
\subsection{Massive Vector Fields\label{s=1,m>0}}

We consider a vector space 
$
{\cal H}
$
of Fock type generated by the vector field 
$
A_{\mu},
$ 
the scalar field 
$
\Phi
$
(with Bose statistics) and the scalar fields 
$
u, \tilde{u}
$
(with Fermi statistics). We suppose that all these (quantum) fields are of mass
$
m > 0.
$
In this vector space we can define a sesquilinear form 
$<\cdot,\cdot>$
in the following way: the (non-zero) $2$-point functions are by definition:
\bea
<\Omega, A_{\mu}(x_{1}) A_{\mu}(x_{2})\Omega> =i~\eta_{\mu\nu}~D_{m}^{(+)}(x_{1} - x_{2}),
\quad
<\Omega, \Phi(x_{1}) \Phi(x_{2})\Omega> =- i~D_{m}^{(+)}(x_{1} - x_{2})
\nonumber \\
<\Omega, u(x_{1}) \tilde{u}(x_{2})\Omega> =- i~D_{m}^{(+)}(x_{1} - x_{2}),
\qquad
<\Omega, \tilde{u}(x_{1}) u(x_{2})\Omega> =- i~D_{m}^{(+)}(x_{1} - x_{2})
\eea
and the $n$-point functions are generated according to Wick theorem. To extend the sesquilinear form to
$
{\cal H}
$
we define the conjugation by
\be
A_{\mu}^{\dagger} = A_{\mu}, \qquad 
u^{\dagger} = u, \qquad
\tilde{u}^{\dagger} = - \tilde{u},
\qquad \Phi^{\dagger} = \Phi.
\ee

Now we introduce
$
{\cal H}
$
the operator $Q$ according to the following formulas:
\bea
~[Q, A_{\mu}] = i~\partial_{\mu}u,\qquad
[Q, u] = 0,\qquad
[Q, \tilde{u}] = - i~(\partial_{\mu}A^{\mu} + m~\Phi)
\qquad
[Q,\Phi] = i~m~u,
\nonumber \\
Q\Omega = 0.
\label{Q-m-1}
\eea
One can prove that $Q$ is well defined. We then have:
\begin{thm}
The operator $Q$ verifies
$
Q^{2} = 0.
$ 
The factor space
$
Ker(Q)/Ran(Q)
$
is isomorphic to the Fock space of particles of mass $m$ and spin $1$ (massive photons, vector bosons).
\end{thm}

\subsection{Gravitational Field\label{s=2,m=0}}

We consider the vector space 
$
{\cal H}
$
of Fock type generated by the symmetric tensor field 
$
h_{\mu\nu}
$ 
(with Bose statistics) and the (ghost) vector fields 
$
u^{\rho}, \tilde{u}^{\sigma}
$
(with Fermi statistics). We suppose that all these (quantum) fields are of null mass. In this vector space we can define a sesquilinear form 
$<\cdot,\cdot>$
in the following way: the (non-zero) $2$-point functions are by definition:
\bea
<\Omega, h_{\mu\nu}(x_{1}) h_{\rho\sigma}(x_{2})\Omega> = - {i\over 2}~
(\eta_{\mu\rho}~\eta_{\nu\sigma} + \eta_{\nu\rho}~\eta_{\mu\sigma}
- \eta_{\mu\nu}~\eta_{\rho\sigma})~D_{0}^{(+)}(x_{1} - x_{2}),
\nonumber \\
<\Omega, u_{\mu}(x_{1}) \tilde{u}_{\nu}(x_{2})\Omega> = i~\eta_{\mu\nu}~
D_{0}^{(+)}(x_{1} - x_{2}),
\nonumber \\
<\Omega, \tilde{u}_{\mu}(x_{1}) u_{\nu}(x_{2})\Omega> = - i~\eta_{\mu\nu}~
D_{0}^{(+)}(x_{1} - x_{2})
\eea
and the $n$-point functions are generated according to Wick theorem. To extend the sesquilinear form to
$
{\cal H}
$
we define the conjugation by
\be
h_{\mu\nu}^{\dagger} = h_{\mu\nu}, \qquad 
u_{\rho}^{\dagger} = u_{\rho}, \qquad
\tilde{u}_{\sigma}^{\dagger} = - \tilde{u}_{\sigma}.
\ee

Now we can introduce the operator $Q$ according to the following formulas:
\bea
~[Q, h_{\mu\nu}] = - {i\over 2}~(\partial_{\mu}u_{\nu} + \partial_{\nu}u_{\mu}
- \eta_{\mu\nu} \partial_{\rho}u^{\rho}),\qquad
[Q, u_{\mu}] = 0,\qquad
[Q, \tilde{u}_{\mu}] = i~\partial^{\nu}h_{\mu\nu}
\nonumber \\
Q\Omega = 0
\label{Q-0-2}
\eea
where by 
$
[\cdot,\cdot]
$
we mean the graded commutator. One can prove that $Q$ is well defined. Indeed, we have the causal commutation relations 
\bea
~[h_{\mu\nu}(x_{1}), h_{\rho\sigma}(x_{2}) ] = - {i\over 2}~
(\eta_{\mu\rho}~\eta_{\nu\sigma} + \eta_{\nu\rho}~\eta_{\mu\sigma}
- \eta_{\mu\nu}~\eta_{\rho\sigma})~D_{0}(x_{1} - x_{2})~\cdot I,
\nonumber \\
~[u(x_{1}), \tilde{u}(x_{2})] = i~\eta_{\mu\nu}~D_{0}(x_{1} - x_{2})~\cdot I
\eea
and the other commutators are null. The operator $Q$ should leave invariant these relations, in particular 
\be
[Q, [ h_{\mu\nu}(x_{1}),\tilde{u}_{\sigma}(x_{2})]] + {\rm cyclic~permutations} = 0
\ee
which is true according to the previous definition. The usefullness of this construction follows from the following result \cite{gravity}:
\begin{thm}
The operator $Q$ verifies
$
Q^{2} = 0.
$ 
The factor space
$
Ker(Q)/Ran(Q)
$
is isomorphic to the Fock space of particles of zero mass and helicity $2$ (gravitons). 
\label{fock-0}
\end{thm}

\subsection{Massive Gravity\label{s=2,m>0}}

We consider a vector space 
$
{\cal H}
$
of Fock type generated (in the sense of Botchers' theorem) by the tensor field 
$
h_{\mu\nu},
$ 
the vector field 
$
v_{\mu}
$
(with Bose statistics) and the vector fields 
$
u_{\mu}, \tilde{u}_{\mu}
$
(with Fermi statistics). We suppose that all these (quantum) fields are of mass
$
m > 0.
$
In this vector space we can define a sesquilinear form 
$<\cdot,\cdot>$
in the following way: the (non-zero) $2$-point functions are by definition:
\bea
<\Omega, h_{\mu\nu}(x_{1}) h_{\rho\sigma}(x_{2})\Omega> = - {i\over 2}~
(\eta_{\mu\rho}~\eta_{\nu\sigma} + \eta_{\nu\rho}~\eta_{\mu\sigma}
- \eta_{\mu\nu}~\eta_{\rho\sigma})~D_{m}^{(+)}(x_{1} - x_{2}),
\nonumber \\
<\Omega, u_{\mu}(x_{1}) \tilde{u}_{\nu}(x_{2})\Omega> = i~\eta_{\mu\nu}~
D_{m}^{(+)}(x_{1} - x_{2}),
\nonumber \\
<\Omega, \tilde{u}_{\mu}(x_{1}) u_{\nu}(x_{2})\Omega> = - i~\eta_{\mu\nu}~
D_{m}^{(+)}(x_{1} - x_{2}),
\nonumber \\
<\Omega, v_{\mu}(x_{1}) v_{\mu}(x_{2})\Omega> =i~\eta_{\mu\nu}~D_{m}^{(+)}(x_{1} - x_{2})
\eea
and the $n$-point functions are generated according to Wick theorem. To extend the sesquilinear form to
$
{\cal H}
$
we define the conjugation by
\bea
h_{\mu\nu}^{\dagger} = h_{\mu\nu}, \qquad 
u_{\rho}^{\dagger} = u_{\rho}, \qquad
\tilde{u}_{\sigma}^{\dagger} = - \tilde{u}_{\sigma}, \qquad
v_{\mu}^{\dagger} = v_{\mu}.
\eea

Now we can introduce the operator $Q$ according to the following formulas:
\bea
~[Q, h_{\mu\nu}] = - {i\over 2}~(\partial_{\mu}u_{\nu} + \partial_{\nu}u_{\mu}
- \eta_{\mu\nu} \partial_{\rho}u^{\rho}),
\nonumber \\
~[Q, u_{\mu}] = 0,\qquad
[Q, \tilde{u}_{\mu}] = i~(\partial^{\nu}h_{\mu\nu} + m v_{\mu}),
\nonumber \\
~[Q, v_{\mu}] = - {i~m\over 2}~u_{\mu}
\nonumber \\
Q\Omega = 0.
\label{Q-m-2}
\eea
One can prove that $Q$ is well defined. Indeed, we have the causal commutation relations 
\bea
~[h_{\mu\nu}(x_{1}), h_{\rho\sigma}(x_{2}) ] = - {i\over 2}~
(\eta_{\mu\rho}~\eta_{\nu\sigma} + \eta_{\nu\rho}~\eta_{\mu\sigma}
- \eta_{\mu\nu}~\eta_{\rho\sigma})~D_{m}(x_{1} - x_{2})~\cdot I,
\nonumber \\
~[u(x_{1}), \tilde{u}(x_{2})] = i~\eta_{\mu\nu}~D_{m}(x_{1} - x_{2})~\cdot I
\nonumber \\
~[v_{\mu}(x_{1}) v_{\mu}(x_{2})] = i~\eta_{\mu\nu}~D_{m}(x_{1} - x_{2})~\cdot I
\eea
and the other commutators are null. The operator $Q$ should leave invariant these relations, in particular 
\bea
[Q, [ h_{\mu\nu}(x_{1}),\tilde{u}_{\sigma}(x_{2})]] + {\rm cyclic~permutations} = 0,
\nonumber \\
~[Q, [ v_{\mu}(x_{1}),\tilde{u}_{\sigma}(x_{2})]] + {\rm cyclic~permutations} = 0.
\eea

We have the result \cite{massive}:
\begin{thm}
The operator $Q$ verifies
$
Q^{2} = 0.
$ 
The factor space
$
Ker(Q)/Ran(Q)
$
is isomorphic to the Fock space of particles of mass $m$ and spin $2$ (massive gravitons) plus a spin $0$ particle of mass $m$.
\end{thm}

\subsection{The General Case}

The situations described above are susceptible to the following generalizations. First we consider the Yang-Mills case. We take a system of 
$
r_{1}
$ 
species of particles of null mass and helicity $1$, that means we use in the first part of this Section 
$
r_{1}
$ 
triplets
$
(A^{\mu}_{a}, u_{a}, \tilde{u}_{a}), a \in I_{1}
$
of massless fields; here
$
I_{1}
$
is a set of indices of cardinal 
$
r_{1}.
$
All the relations have to be modified by appending an index $a$ to all these fields. In the massive case we have to consider 
$
r_{2}
$ 
quadruples
$
(A^{\mu}_{a}, u_{a}, \tilde{u}_{a}, \Phi_{a}),  a \in I_{2}
$
of fields of mass 
$
m_{a}
$ 
as in section \ref{s=1,m>0}; here
$
I_{2}
$
is a set of indices of cardinal 
$
r_{2}.
$
We want to include some arbitrary scalar fields with indices
$
a \in I_{3}.
$
Then we take 
$
I = I_{1} \cup I_{2} \cup I_{3}
$
a set of indices and for any index we take a quadruple
$
(A^{\mu}_{a}, u_{a}, \tilde{u}_{a},\Phi_{a}), a \in I
$
of fields with the following conventions:
(a) the first entry are vector fields and the last three ones are scalar fields;
(b) the fields
$
A^{\mu}_{a},~\Phi_{a}
$
are obeying Bose statistics and the fields
$
u_{a},~\tilde{u}_{a}
$
are obeying Fermi statistics;
(c) For
$
a \in I_{1}
$
we impose 
$
\Phi_{a} = 0
$
and we take the masses to be null
$
m_{a} = 0;
$
(d) For
$
a \in I_{2}
$
we take all the masses strictly positive:
$
m_{a} > 0;
$
(e) For 
$
a \in I_{3}
$
we take 
$
A_{a}^{\mu}, u_{a}, \tilde{u}_{a}
$
to be null and the fields
$
\Phi_{a} \equiv \phi^{H}_{a} 
$
of mass 
$
m^{H}_{a} \geq 0.
$
The fields
$
u_{a},~\tilde{u}_{a},~~a \in I_{1} \cup I_{2}
$
and
$
\Phi_{a}~~a \in I_{2}
$
are called {\it ghost fields} and the fields
$
\phi^{H}_{a},~~a \in I_{3} 
$
are called {\it Higgs fields};
(f) we include spinorial matter fields also i.e some set of Dirac fields with Fermi statistics:
$
\Psi_{A}, A \in I_{4};
$ 
(g) we consider that the Hilbert space is generated by all these fields applied on the vacuum and  define in 
$
{\cal H}
$
the gauge charge operator $Q$ according to the following formulas for all indices
$
a \in I:
$
\bea
~[Q, v^{\mu}_{a}] = i~\partial^{\mu}u_{a},\qquad
[Q, u_{a}] = 0,
\nonumber \\
~[Q, \tilde{u}_{a}] = - i~(\partial_{\mu}v^{\mu}_{a} + m_{a}~\Phi_{a})
\qquad
[Q,\Phi_{a}] = i~m_{a}~u_{a},
\label{Q-general}
\eea
\be
[Q,\Psi_{A}] = 0,
\ee
and
\be
Q\Omega = 0.
\ee
If we want to include gravitons also then we extend the Fock space including the corresponding free fields 
$
h_{\mu\nu}, u^{\rho}, \tilde{u}^{\sigma}
$
(for massless gravity) or
$
h_{\mu\nu}, u^{\rho}, \tilde{u}^{\sigma}, v^{\lambda}
$
(for massive gravity) and we extend the definition of the gauge charge $Q$ in a natural way using (\ref{Q-0-2}) or (\ref{Q-m-2}). In this way the Fock space will describe a system of Yang-Mills particles together with (massless or massive) gravitons.
\section{The Cohomology of the Operator $d_{Q}$\label{q}}

We know that the condition 
$
[Q, T] = i~\partial_{\mu}T^{\mu}
$
means that the expression $T$ leaves invariant the physical Hilbert space (at least in the adiabatic limit).

Now we have the physical justification for solving the cohomology problem namely to determine the cohomology of the operator 
$
d_{Q} = [Q,\cdot]
$
induced by $Q$ in the space of Wick polynomials. One can solve this problem in a quite general setting using the jet bundle formalism \cite{cohomology} and \cite{cohomology2}. However for all practical purposes we need only a very particular case. Let us denote by 
$
{\cal P}^{{\rm tri},5}
$
the set of Wick polynomials in the fields and their derivatives subject to the following restrictions: if
$
T \in {\cal P}^{{\rm tri},5}
$
then $T$ is tri-linear in the fields (and their derivatives) and
$
\omega(T) \leq 5.
$
We are interested in the cohomology of the operator
$
d_{Q}
$
acting in
$
{\cal P}^{{\rm tri},5}.
$
We will denote by
$
Z^{{\rm tri},5}_{Q}
$
and 
$
B^{{\rm tri},5}_{Q}
$
the co-cycles and the co-boundaries of this operator, respectively.

We introduce some notations. Basically we are looking for gauge-invariant variables.

For the case of massless or massive spin $1$ fields we define the {\it field strength} according to
\be
F^{\mu\nu} \equiv \partial^{\mu}A^{\nu} - \partial^{\nu}A^{\mu}
\ee 
and observe that 
\bea
d_{Q}F^{\mu\nu} = 0,
\nonumber \\
\partial_{\rho}F_{\mu\nu} + \partial_{\mu}F_{\nu\rho} + \partial_{\nu}F_{\rho\mu} = 0
\eea
the last relation being called {\it Bianchi identity} (or homogeneous Maxwell equation by physicists). We denote by
$
F^{(0)}_{\mu\nu;\rho}
$
the traceless part (in all indices) of the expression
$
\partial_{\rho}F_{\mu\nu}
$
(which also verifies the Bianchi identities).

In the case of a massive vector field it is convenient to introduce another notation, namely:
\be
\phi_{\mu} \equiv \partial_{\mu}\Phi - m~A_{\mu}
\ee
and we observe that
\be
d_{Q}\phi_{\mu} = 0.
\ee
We denote by 
$
\phi^{(0)}_{\mu\nu}
$
the traceless part of the expression
\be
\partial_{\mu}\partial_{\nu}\Phi - {1\over 2}~m~(\partial_{\mu}A_{\nu} + \partial_{\nu}A_{\mu}).
\ee

For the general Yang-Mills system we have to append an index
$
a \in I_{1} \cup I_{2}
$
and we denote by
$
{\cal P}^{{\rm tri},5}_{0} \subset {\cal P}^{{\rm tri},5}
$
the space of polynomials in the variables
$
u_{a},~F_{a\mu\nu},~F^{(0)}_{a\mu\nu;\rho}~(a \in I_{1})
$
and
$
F_{a\mu\nu},~F^{(0)}_{a\mu\nu;\rho},~\phi_{a\mu},~\phi^{(0)}_{a\mu\nu}~(a \in I_{2}).
$ 
Then we have the following result \cite{cohomology}:
\begin{thm}
Let 
$
p \in Z^{{\rm tri},5}_{Q}.
$
Then $p$ is cohomologous to a polynomial of the form 
$
p = p_{1} + d_{Q}p_{2}
$ 
where
$
p_{1} \in {\cal P}^{{\rm tri},5}_{0}
$
and
$
p_{2} \in {\cal P}^{{\rm tri},5}.
$ 
\label{q1}
\end{thm}

The expressions of the type 
$
d_{Q}p_{2}
$ 
do appear because for massive Yang-Mills fields it is possible that the gauge charge operator does not raise the canonical dimension (for instance this is true for a monomial having a factor
$
\Phi_{a}
$)

In the case of the gravitational field it is also convenient to introduce some other notations: first
\be
h \equiv \eta^{\mu\nu}h_{\mu\nu} \qquad
\hat{h}_{\mu\nu} \equiv h_{\mu\nu} - {1\over 2}~\eta_{\mu\nu}~h
\label{h}
\ee
and the we define the {\it Christoffel symbols} according to:
\be
\Gamma_{\mu;\nu\rho} \equiv \partial_{\rho}\hat{h}_{\mu\nu} + \partial_{\nu}\hat{h}_{\mu\rho} - \partial_{\mu}\hat{h}_{\nu\rho}.
\ee 

The expression
\be
R_{\mu\nu;\rho\sigma} \equiv \partial_{\rho}\Gamma_{\mu;\nu\sigma} 
- (\rho \leftrightarrow \sigma)
\ee
is called the {\it Riemann tensor}; we can easily prove 
\bea
R_{\mu\nu;\rho\sigma} = - R_{\nu\mu;\rho\sigma} = - R_{\mu\nu;\sigma\rho} = R_{\rho\sigma;\mu\nu}, 
\nonumber \\
d_{Q}R_{\mu\nu;\rho\sigma} = 0,
\nonumber \\
R_{\mu\nu;\rho\sigma} + R_{\mu\rho;\nu\sigma} + R_{\mu\sigma;\nu\rho} = 0;
\nonumber \\
d_{\lambda}R_{\mu\nu;\rho\sigma} + d_{\rho}R_{\mu\nu;\sigma\lambda} + d_{\sigma}R_{\mu\nu;\lambda\rho} = 0
\eea
the last two relations are called {\it Bianchi identities}. 

We denote by
$
R^{(0)}_{\mu\nu;\rho\sigma;\lambda}.
$
the traceless part in all indices of
$
\partial_{\lambda}R^{(0)}_{\mu\nu;\rho\sigma}.
$
and we also define
\be
u_{\mu\nu} = u_{[\mu\nu]} \equiv {1\over 2}~(\partial_{\mu}u_{\nu} - \partial_{\nu}u_{\mu}).
\ee

Now we have the following result \cite{cohomology2}:
\begin{thm}
Let 
$
p \in Z^{{\rm tri},5}_{Q}.
$
Then $p$ is cohomologous to a polynomial in 
$u_{\mu}, u_{\mu\nu}$ 
and in
$
R^{(0)}_{\mu\nu;\rho\sigma},~R^{(0)}_{\mu\nu;\rho\sigma;\lambda}.
$
\label{q20}
\end{thm}

In the case of massive gravity we also define the expressions
\bea
\varphi_{\mu\nu} \equiv 
\partial_{\mu}v_{\nu} + \partial_{\nu}v_{\mu} - \eta_{\mu\nu} \partial_{\rho}v^{\rho} - m~h_{\mu\nu}
\nonumber \\
\varphi \equiv \eta^{\mu\nu}~\varphi_{\mu\nu}
\eea
and observe that we also have
\be
d_{Q}\varphi_{\mu\nu} = 0.
\ee
By
$
\varphi^{(0)}_{\mu\nu;\rho}
$
we denote the traceless part of
$
\partial_{\rho}\varphi_{\mu\nu}.
$

As in the Yang-Mills case we denote by
$
{\cal P}^{{\rm tri},5}_{0} \subset {\cal P}^{{\rm tri},5}
$
the space of polynomials in the variables
$
R^{(0)}_{\mu\nu;\rho\sigma},~R^{(0)}_{\mu\nu;\rho\sigma;\lambda}
$
and
$
\varphi_{\mu\nu},~\varphi,~\varphi^{(0)}_{\mu\nu;\rho},~\partial_{\rho}\varphi. 
$ 
Then we have \cite{cohomology2}:
\begin{thm}
Let 
$
p \in Z^{{\rm tri},5}_{Q}.
$
Then $p$ is cohomologous to a polynomial of the form 
$
p = p_{1} + d_{Q}p_{2}
$ 
where
$
p_{1} \in {\cal P}^{{\rm tri},5}_{0}
$
and
$
p_{2} \in {\cal P}^{{\rm tri},5}.
$ 
\label{q2m}
\end{thm}

We can obtain from the theorems above the description of the co-cycles in the general case of Yang-Mills fields interacting with massless or massive gravity using K\"unneth theorem \cite{Dr}. This means that we must consider polynomials in invariants of both Yang-Mills and gravity type.

We mention in closing that for polynomials from 
$
{\cal P}^{{\rm tri},5}
$
Poincar\'e lemma holds i.e. we have for any
$
 T^{\mu} \in {\cal P}^{{\rm tri},5} 
$
\be
\partial_{\mu}T^{\mu} = 0 \qquad \Longrightarrow \qquad T^{\mu} = \partial_{\nu}T^{\mu\nu}
\ee
where the expression
$
T^{\mu\nu} \in {\cal P}^{{\rm tri},5}
$
is antisymmetric \cite{cohomology}.
\section{The Interaction of Gravity with other Quantum Fields\label{int1}}

We here consider a system of massive and massless Yang-Mills fields
$
(v^{\mu}_{a}, u_{a}, \tilde{u}_{a}, \Phi_{a}),  a \in I
$
and we determine the coupling with the massless gravitational field
$
h_{\mu\nu},~u^{\rho},~\tilde{u}^{\sigma}.
$ 
By definition the ghost number is the sum of the ghost numbers of the YM and gravity sectors. As in \cite{cohomology2} we consider that the interaction has null ghost number and its canonical dimension is bounded by $5$. We will get an expression of the form
\be
T_{\rm int} = T^{\rm YM}_{\rm int} + T^{\rm scalar}_{\rm int} + T^{\rm Fermi}_{\rm int}
\ee
with the scalar and Fermi contributions being the same as in \cite{cohomology2}. We concentrate only on the Yang-Mills contribution and we have our main result. We use the following definitions. A Wick polynomial $T$ is called a {\it relatively co-cycle} {\it iff} it verifies the relation
$
d_{Q}T = \partial_{\mu}T^{\mu};
$
two Wick polynomials are {\it relatively cohomologous} {\it iff} the differ by an expression of the type
$
d_{Q}B + \partial_{\mu}B^{\mu}.
$
As it was already seen in previous papers, one can reduce the problem of determining the relative cohomology groups to the cohomology of the operator
$
d_{Q}
$
using the descent procedure.
\begin{thm}
The expression
$
T^{\rm YM}_{\rm int}
$
is relatively cohomologous to
\bea
t^{\rm YM}_{\rm int} \equiv \sum_{a,b \in I_{1}}~f_{ab}~
(4 h_{\mu\nu}~F_{a}^{\mu\rho}~{F_{b}^{\nu}}_{\rho} 
- h~F_{a\rho\sigma}~F_{b}^{\rho\sigma} 
+ 4~u_{\mu}~d_{\nu}\tilde{u}_{a}~F_{b}^{\mu\nu})
\nonumber \\
+ \sum_{a,b \in I_{2}}~f_{ab}~
(4 h_{\mu\nu}~F_{a}^{\mu\rho}~{F_{b}^{\nu}}_{\rho} 
- h~F_{a\rho\sigma}~F_{b}^{\rho\sigma} 
+ 4~u_{\mu}~d_{\nu}\tilde{u}_{a}~F_{b}^{\mu\nu}
- 4~h_{\mu\nu}~\phi_{a}^{\mu}~\phi_{b}^{\nu}
- 4~m_{a}~u_{\mu}~\tilde{u}_{a}~\phi_{b}^{\mu})
\label{t-int}
\eea
with the constants
$
f_{ab}
$
symmetric
$
f_{ab} = f_{ba}
$
and real.

(ii) The relation 
$
d_{Q}t^{\rm YM}_{\rm int} = i~d_{\mu}t^{{\rm YM},\mu}_{\rm int}
$
is verified by:
\bea
t_{\rm int}^{{\rm YM},\mu} \equiv \sum_{a,b \in I_{1}}~f_{ab}~
( u^{\mu}~F_{a}^{\rho\sigma}~F_{b\rho\sigma} 
+ 4~u^{\rho}~F_{a}^{\mu\nu}~F_{b\nu\rho} )
\nonumber \\
+ \sum_{a,b \in I_{2}}~f_{ab}~
( u^{\mu}~F_{a}^{\rho\sigma}~F_{b\rho\sigma} 
+ 4~u^{\rho}~F_{a}^{\mu\nu}~F_{b\nu\rho}
- 2~u^{\mu}~\phi_{a\nu}~\phi_{b}^{\nu} 
+ 4~m_{a}~u_{\nu}~\phi_{a}^{\mu}~\phi_{b}^{\nu})
\eea
and we also have
\be
d_{Q}t_{\rm int}^{{\rm YM},\mu} = 0.
\ee
\label{T-int}
\end{thm}
{\bf Proof:} (i) By hypothesis we have
\be
d_{Q}T_{\rm int} = i~\partial_{\mu}T_{\rm int}^{\mu}
\label{descent-tint}
\ee
and the descent procedure based on Poincar\'e lemma (see \cite{cohomology}, \cite{cohomology2} and \cite{descent}) leads to
\bea
d_{Q}T_{\rm int}^{\mu} = i~\partial_{\nu}T_{\rm int}^{\mu\nu}.
\nonumber\\
d_{Q}T_{\rm int}^{\mu\nu} = i~\partial_{\rho}T_{\rm int}^{\mu\nu\rho}
\nonumber \\
d_{Q}T_{\rm int}^{\mu\nu\rho} = i~\partial_{\sigma}T_{\rm int}^{\mu\nu\rho\sigma}
\nonumber \\
d_{Q}T_{\rm int}^{\mu\nu\rho\sigma} = 0
\label{descent-T-int}
\eea
and can choose the expressions
$
T_{\rm int}^{I}
$
to be Lorentz covariant; we also have
\be
gh(T_{\rm int}^{I}) = |I|, \omega(T_{\rm int}^{I}) \leq 5. 
\ee 

From the last relation and the theorems \ref{q1} and \ref{q20} we find that
\be
T_{\rm int}^{\mu\nu\rho\sigma} = d_{Q}B^{\mu\nu\rho\sigma} 
+ T_{{\rm int},0}^{\mu\nu\rho\sigma}
\ee
with
$
T_{{\rm int},0}^{\mu\nu\rho\sigma}
$
a polynomial in the invariants described in theorems \ref{q1} and \ref{q20} and we can choose the expressions
$
B^{\mu\nu\rho\sigma}
$
and
$
T_{{\rm int},0}^{\mu\nu\rho\sigma}
$
completely antisymmetric. The generic form of 
$
T_{{\rm int},0}^{\mu\nu\rho\sigma}
$
can be easily obtained. If we substitute the expression of
$
T_{\rm int}^{\mu\nu\rho\sigma}
$
in the third relation (\ref{descent-T-int}) we find out
\be
d_{Q}(T_{\rm int}^{\mu\nu\rho} - i~\partial_{\sigma}B^{\mu\nu\rho\sigma}) 
= i~\partial_{\sigma}T_{{\rm int},0}^{\mu\nu\rho\sigma}
\ee 
so the expression in the right hand side must be a co-boundary and we immediately obtain
$
T_{{\rm int},0}^{\mu\nu\rho\sigma} = 0
$
so:
\be
T_{\rm int}^{\mu\nu\rho\sigma} = d_{Q}B^{\mu\nu\rho\sigma}
\ee
and
\be
d_{Q}(T_{\rm int}^{\mu\nu\rho} - i~\partial_{\sigma}B^{\mu\nu\rho\sigma}) = 0.
\ee 
We continue in the same way and obtain:
\be
T_{\rm int}^{\mu\nu\rho} = d_{Q}B^{\mu\nu\rho} + i~\partial_{\sigma}B^{\mu\nu\rho\sigma}
\ee
and
\be
T_{\rm int}^{\mu\nu} = d_{Q}B^{\mu\nu} + i~\partial_{\rho}B^{\mu\nu\rho}.
\ee

(ii) We substitute the expression of
$
T_{\rm int}^{\mu\nu}
$
in the first relation (\ref{descent-T-int}) and get:
\be
d_{Q}(T_{\rm int}^{\mu} - i~\partial_{\nu}B^{\mu\nu}) = 0
\ee

Now it is again time to use theorems \ref{q1} and \ref{q20} to obtain
\be
T_{\rm int}^{\mu} = d_{Q}B^{\mu} + i~\partial_{\nu}B^{\mu\nu} + T^{\mu}_{{\rm int},0}
\ee 
where
$
T_{{\rm int},0}^{\mu}
$ 
is a polynomial in the invariants appearing in the theorems \ref{q1} and \ref{q20}. We get from the first relation (\ref{descent-T-int}) 
\be
d_{Q}(T_{\rm int} - i~\partial_{\mu}B^{\mu}) = i~\partial_{\mu}T^{\mu}_{{\rm int},0}
\ee
so the right hand side must be a co-boundary. At this stage of the computation some non-trivial co-cycles do appear in
$
T_{{\rm int},0}^{\mu}
$ 
We consider only the Yang-Mills sector; then we have:
\bea
T^{{\rm YM},\mu}_{{\rm int},0} = f_{ab}^{(1)}~u^{\mu}~F_{a}^{\rho\sigma}~F_{b\rho\sigma} 
+ f_{ab}^{(2)}~u^{\rho}~F_{a}^{\mu\nu}~F_{b\nu\rho}
\nonumber \\
g_{ab}^{(1)}~u^{\mu}~\phi_{a\nu}~\phi_{b}^{\nu} 
+ g_{ab}^{(2)}~u_{\nu}~\phi_{a}^{\mu}~\phi_{b}^{\nu} + \cdots
\eea
where by 
$
\cdots
$ 
we mean other terms; we can impose the symmetry conditions
\be
f_{ab}^{(1)} = a \leftrightarrow b, \qquad g_{ab}^{(1)} = a \leftrightarrow b.
\ee
If one computes the divergence 
$
\partial_{\mu}T^{{\rm YM},\mu}_{{\rm int},0}
$
and imposes the condition that it is a co-boundary, then one gets:
\bea
f_{ab}^{(2)} = 4~f_{ab}^{(2)}, \qquad g_{ab}^{(2)} = - 2~g_{ab}^{(1)},
\nonumber \\
2~f_{ab}^{(1)} + g_{ab}^{(1)} = 0~(\forall a,b \in I_{2}),\qquad
f_{ab}^{(1)} = 0~(\forall a \in I_{1}, b \in I_{2})
\eea
and also the terms $\cdots$ from the expression of
$
T^{{\rm YM},\mu}_{{\rm int},0}
$
are null. The first equality above comes from terms without a mass factor and the second one from terms with a mass factor. It follows that one can take
\be
T_{{\rm int},0}^{{\rm YM},\mu} = t^{{\rm YM},\mu}_{\rm int}
\ee
with
$
t^{{\rm YM},\mu}_{\rm int}
$
the expression from the statement of the theorem. Because we have by direct computation
$
d_{Q}t^{\rm YM}_{\rm int} = i~\partial_{\mu}t_{\rm int}^{{\rm YM},\mu}
$
we get
\be
d_{Q}(T_{\rm int} - t^{\rm YM}_{\rm int} - i~\partial_{\mu}B^{\mu}) = 0
\ee
so known results lead to
\be
T_{\rm int} = t^{\rm YM}_{\rm int} + d_{Q}B + i~\partial_{\mu}B^{\mu} + T_{{\rm int},0}
\ee 
where
$
T_{{\rm int},0}
$
is a polynomial in the invariants appearing in the theorems \ref{q1} and \ref{q20}. But there are no such expression i.e.
$
T_{{\rm int},0} = 0
$
and we have 
\be
T_{\rm int} = t^{\rm YM}_{\rm int} + d_{Q}B + i~\partial_{\mu}B^{\mu}
\ee 
which is the final result.

We point out that in all these computations one should consider all polynomials in the invariants, even or odd with respect to parity invariance. Indeed, because the Yang-Mills interaction is not parity invariant, there are no reasons to suppose that the interaction with gravity is parity invariant. Fortunately, the odd sectors do not produce non-trivial obstructions to the descent procedure.
$\qed$

Let us note that the result of the theorem stays true if we replace massless gravity by massive gravity. In the proof we will have to use now theorems \ref{q1} and \ref{q2m}.

We close by mentioning that the interaction between massive Yang-Mills fields and gravity (i.e. the second line in the formula (\ref{t-int}) can be put into a simpler form. For simplicity we first consider one massive vector field i.e. 
$
|I_{2}| = 1.
$ 
In this case we can skip the index 
$a = 1$
and we define the {\it physical part of the vector field }
$
A_{\mu}
$
according to the formula \cite{Sc2}:
\be
A^{\rm phys}_{\mu} \equiv A_{\mu} + {1\over m^{2}}~\partial^{\mu}~\partial_{\nu}~A_{\nu}.
\ee
This field has the following properties:
\be
d_{Q}A^{\rm phys}_{\mu} = 0, \qquad \partial^{\mu}A^{\rm phys}_{\mu} = 0.
\ee
Then one can prove by some computations the following formula:
\be
h_{\mu\nu}~\phi_{a}^{\mu}~\phi_{b}^{\nu}
+ m_{a}~u_{\mu}~\tilde{u}_{a}~\phi_{b}^{\mu} 
- u_{\mu}~d_{\nu}\tilde{u}_{a}~F_{b}^{\mu\nu}
= m^{2}~h^{\mu\nu}~A^{\rm phys}_{\mu}~A^{\rm phys}_{\nu} + d_{Q}B + \partial_{\mu}B^{\mu}
\ee
i.e we can express the interaction between the massive vector field
$
A_{\mu}
$
and gravity in terms of the physical part of
$
A_{\mu}
$
in a standard form:
\be
t_{\rm int} = \hat{h}_{\mu\nu}~{\cal T}^{\mu\nu}
\label{t-int-phys}
\ee
where we have defined $h$ and 
$
\hat{h}_{\mu\nu}
$
in formula (\ref{h}) and
\be
{\cal T}_{\mu\nu} \equiv F_{\mu\rho}~{F_{\nu}}^{\rho} 
- {1\over 4}~\eta_{\mu\nu}~F_{\rho\sigma}~F^{\rho\sigma} 
- m^{2}~A^{\rm phys}_{\mu}~A^{\rm phys}_{\nu}
+ {m^{2}\over 2}~\eta_{\mu\nu}~A^{\rm phys,\rho}~A^{\rm phys}_{\rho}
\ee
is the {\it energy-momentum tensor}; one can prove directly that it is conserved:
\be
\partial_{\nu}~{\cal T}^{\mu\nu} = 0.
\ee 

In this form gravity couples to the physical degrees of freedom only, because there is no coupling to the ghost and 
$
A^{\rm phys}_{\mu}
$
contains the three transverse physical modes only [11]. However, the new expression of the interaction Lagrangian has canonical dimension $7$ so for the purpose of perturbation theory it is better to work with the expression appearing in the theorem. 

Note that the ghost couplings like
$
u_{\mu}~d_{\nu}\tilde{u}_{a}~F_{b}^{\mu\nu}
$
do not contribute to the $S$-matrix elements between physical states in arbitrary order. Nevertheless, these couplings are necessary for gauge invariance of the theory. On the other hand the pure Yang-Mills couplings like
$
f_{abc}~u_{a}~A_{b}^{\mu}~\partial_{\mu}\tilde{u}_{c}
$
do contribute to physical $S$-matrix elements in ghost-anti-ghost loops.

If one considers that the tensor
$
f_{ab}
$
in the formula (\ref{t-int}) is diagonal then the energy-momentum tensor is additive with respect to the vector fields and the formula (\ref{t-int-phys}) extends naturally. (This property of the tensor
$
f_{ab}
$
follows from gauge invariance in the second order of the perturbation theory; this will be done in another publication). 
\section{Conclusions}
In classical general relativity one says that gravity couples to everything which carries energy and momentum. That means one must have an energy-momentum tensor which is conserved
$
\partial_{\nu}~{\cal T}^{\mu\nu} = 0. 
$
In quantum theory such a tensor does not exist in general. For example, in massless Yang-Mills theory and even in quantum electro-dynamics the naive (free) energy-momentum tensor 
$
{\cal T}^{\mu\nu} \equiv F^{\mu\rho}~{F^{\nu}}_{\rho} 
- {1\over 4}~\eta^{\mu\nu}~~F_{\rho\sigma}~F^{\rho\sigma}
$
is not conserved because
$
\partial_{\mu}~A^{\mu} = 0 
$
cannot be assumed as an operator equation. In this situation it is hard to find the correct gravitational couplings by classical Lagrangian arguments.

Fortunately, the requirement of gauge invariance
$
d_{Q}T = i~\partial_{\mu}T^{\mu}
$
is so strong that it determines all couplings uniquely, if some natural additional properties are assumed. It is a surprise that the resulting gravitational couplings contain ghost fields because they do not ``carry energy and momentum". The paradox is resolved by observing that the ghost coupling terms do not contribute to $S$-matrix elements between physical states. 

\end{document}